\documentclass{icrc2009}

\usepackage{graphicx}   
\usepackage{caption}    
\usepackage[font=footnotesize]{subfig} 
\usepackage{fixltx2e}
\usepackage{url}

\newcommand{\shorttitle}[1]%
{\markboth{Proceedings of the 31\MakeLowercase{$^{st}$} ICRC, {\L}\'{o}d\'{z} 2009}{#1} }

\renewcommand{\theenumi}{\roman{enumi}.}

\begin{document}
\title{An unbinned test for Quantum Gravity effects in high-energy light-curves.}

\author{\IEEEauthorblockN{Ulisses Barres de
Almeida and
              Michael Daniel}
                            \\
\IEEEauthorblockA{Department of Physics, University of Durham,
England.}}

\shorttitle{Barres de Almeida \& Daniel - Test for Quantum Gravity}
\maketitle

\begin{abstract}
Some models of quantum gravity can predict observable effects on the
propagation of light: most notably an energy dependent dispersion,
where the speed of light is seen to vary with the energy of the
photon. As quantum gravity effects should appear at the Planck scale
they will be very small and so require very high energy photons to
travel large distances before even becoming noticeable. Precisely
because this effect is greater for the most energetic photons (dt
$\sim$ 10 s/TeV/Gpc), ground-based gamma-ray measurements of large
AGN flares are the ideal resource for performing such tests. The
modest photon flux combined with the fact that these experiments are
capable of recording the photon times with great resolution suggests
the use of unbinned algorithms as an optimal solution for testing
models of quantum gravity. In this paper we discuss the application
of a non-parametric test to such datasets, analysing its limitations
and exploring the potential benefits.
  \end{abstract}

\begin{IEEEkeywords}
 Lorentz invariance violation, statistical methods, quantum-gravity
\end{IEEEkeywords}

\section{Introduction}
The study of phenomena in which both quantum mechanical and general
relativistic effects are important, motivated theoretical efforts to
construct a theory capable of describing gravitation at the
subatomic level: the so-called Quantum Gravity (QG) theories. One of
the most fundamental results, common to several competing approaches
to QG, is the quantization of the space-time continuum, which
appears in the form of a space-time uncertainty relation $\Delta x
\Delta t \ge const$ (e.g., \cite{ellisb}).


A consequence of this discreteness of space-time is that the vacuum
will interact with energetic photons on the Planck scale, acting
analogously to a medium that absorbs and re-emits radiation by
excitation of its internal degrees of freedom \cite{feynman}.
Lorentz invariance violation (LIV) arises in this context due to a
modified dispersion relation for the photon, resulting from a
non-trivial, spectral dependent refractive index for the vacuum of
the form $n - 1 \sim E_{\gamma}/E_{QG}$ \footnote{Notice that here,
unlike an ordinary medium, the vacuum refractive index {\it
increases} for smaller wavelengths in QG models.}, where $E_{QG}$ is
the energy-scale for QG, expected to be of the order of the Planck
energy $\sim E_{P}\simeq 10^{19}$ GeV \cite{ellisb}.

For photons of energies $E << E_{QG}$, the perturbed dispersion
relation can be approximated by a series expansion of the form
\cite{amelino-camelia}:

\begin{equation}\label{equ1}
c^{2} {\bf p}^{2}=E^{2}[1+\xi E/E_{QG}+O(E^{2}/E_{QG}^{2})]
\end{equation}

Despite the vanishingly small velocity corrections, of the order of
$10^{-15}c$ for a 1 TeV photon, the observation of extragalactic
gamma-ray sources such as gamma-ray bursts (GRBs) and active
galactic nuclei (AGNs) is a promising laboratory to test this
prediction of QG theories. This is because the variations on the
speed of light, integrated over the large propagation distances of
the photons, result in sizeable delays that could be directly
measured by high-accuracy timing experiments \cite{amelino-camelia},
which including the cosmological effects of propagation on an
expanding universe \cite{ellisa}:


\begin{equation}\label{equ3}
\Delta t = H_{0}^{-1} \frac{\Delta E}{E_{QG}}\int_{0}^{z} h^{-1}dz,
\end{equation}
where $H_{0}$ and $h$ are respectively the Hubble constant and its
associated dimensionless parameter \cite{hinshaw}, and $z$ is the
redshift of the source. In the analysis of broad spectral band
light-curves, this delay will manifest as a time-lag between the
arrival times of the lowest and the highest energy photons of
$\approx 10$$\xi$ s Gpc$^{-1}$ TeV$^{-1}$.

Traditionally, high-energy experiments have drawn from this
principle and, by splitting the light-curves in two or more energy
bins, have looked for significant shifts in the times of bursts or
sharp features between them, deriving upper limits to the magnitude
and energy-scale of the QG effects. In the following section we will
briefly review the current status and results of these searches,
before proceeding to the presentation of our method.

\section{Current Results of Time-lag Measurements}
In recent years, several high-energy experiments have begun to
perform timing analysis in order to identify energy dependent lags
in the light curves of distant sources such as GRBs and AGNs. In
principle, the former would be the preferred targets for the study
because of their large distances and extremely short burst features,
reaching down to sub-second and even millisecond timescales
(\cite{fishman} and \cite{bhat}). The advent of FERMI brings great
prospects to the search for LIV signatures, due to a significant
increase in sensitivity. The most constraining GRB results to date
come from recent FERMI observations of GRB080916C, and give a robust
lower limit of $1.3\times 10^{18} GeV/c^{2}$ to the energy scales of
QG \cite{fermi}.

Nevertheless, AGN observations with ground-based gamma-ray
telescopes carry the advantage of observations at much higher
energies, increasing the magnitude of the QG-induced lags one is
seeking to tens of seconds. Recent results by HESS \cite{qghess} and
MAGIC \cite{qgmagic} provide lower limits for the onset of QG
effects of $1.44\times 10^{18}$ GeV and $0.52\times 10^{18}$ GeV
respectively, in agreement with the newest GRB results. For
specialized reviews of these latest results see \cite{ellisc} and
\cite{wagner}.

\section{Unbinned Tests for the Detection of Photon Dispersion}
Given the discrete nature of the high-energy data, tests that
exploit the full information content of the light-curves by looking
at individual photons are a natural choice to exploit the maximum
 sensitivity of the experiments. The new method we
propose for detecting spectral time-lags in the light-curves of
high-energy sources has its fundamental idea drawn from the original
approaches of \cite{albert} or \cite{scargle}. It consists of using
the linear approximation to the energy-dependent delay given in
equation (2) to apply a systematic correction $\tau > 0$ to the
arrival times of each individual recorded photon of the form:

\begin{equation}\label{equ5}
\Delta t = - \tau E,
\end{equation}
so as to \textit{cancel} any putative QG effects on photon
propagation. Since the applied correction is to be the exact inverse
of the original dispersion, the optimal correction is a direct
estimate of the QG energy scale and dispersion magnitude
$\xi/E_{P}$.

The QG signature is asymmetric, always introducing a dispersion to
the original burst profile. Therefore the correction $\tau$ assumed
to most closely cancel the time-of-flight delays is expected also to
return a burst profile which is maximally sharp, according to an
appropriately chosen measure, so that the problem is then reduced to
that of the maximization of a cost function. In \cite{scargle}, the
two proposed cost functions are the Shannon Information (or an
alternative information-entropy measure) and the average intra-pulse
photon-interval, whereas in \cite{albert} the authors sought to
maximize the total power of the burst around its maximum.

In any approach of this sort there are two basic assumptions in
play, which represent limitations to the method:
\begin{enumerate}
\item all energy-dependent dispersion corrected in the cancelation
algorithm is supposed extrinsic to the source and due to QG, since
we cannot account for effects intrinsic to the emission process;
\item the maximally sharp burst retrieved by the linear correction is
assumed to be an accurate representation of the original burst
profile.
\end{enumerate}

Whilst (i) is an unavoidable condition in (preferential)
non-parametric approaches\footnote{See \cite{martinez} for an
alternative non-binned approach that try to surpass this limitation
by introducing a model-dependent cost function, incorporating
properties of the source emission process.}, we propose a method
that avoids (ii), substituting it by a somewhat less arbitrary
assumption. As discussed in \cite{scargle} the problem with this
latter condition is that it cannot handle equally well cases where
we have to deal with overlapping or asymmetric bursts, in which the
maximum sharpness condition may not lead to the correct dispersion
cancelation parameter.

\section{Kolmogorov distance}
Given two random variables $X$ and $Y$ in $\Re$, the simplest
measure of the \textit{difference} between their probability
distributions (pdfs) is the \textit{Kolmogorov distance}, introduced
by Kolmogorov in 1933 as a metric for random variables in
probability space (see \cite{kolmogorov} and \cite{zolotarev}). For
$F_{X}(x) =$ prob$(X \le x)$ and $F_{Y}(x)=$prob$(Y \le x)$,
cumulative distribution functions (cdfs) of $X$ and $Y$
respectively, the Kolmogorov distance is defined as

\begin{equation}\label{equ6}
D_{K} \equiv  \sup_{x \in \Re}|F_{X}(x) - F_{Y}(x)|,
\end{equation}
the maximum vertical distance between the two cdfs.

Given a broad spectral range light-curve with sufficient photon
statistics, we can meaningfully bin the data in low and high energy
bands, creating two light-curves that should in principle superpose,
provided that the high and low energy photons were produced
simultaneously at the source, without any intrinsic net delay (same
as condition 1 from last paragraph). After propagation, if QG
effects are present, the profiles of any given burst in the light
curve will differ in the two energy bands, due to the different
amount of dispersion of the photons, so that the high energy ones
will be more strongly shifted towards later arrival times.

Following \cite{scargle}, we represent the bursts by a normalised
probability distribution. We construct a photon cell $x_{i} =
1/dt_{i}$ at the place of each photon $i$, where $dt_{i}$ is the
waiting time of each photon and $x_{i}$ is then indicative of the
photon density at each time. We then transform these densities into
normalised probability measures by defining $p_{n}=x_{n}/ \Sigma
x_{n}$ for every cell $n$\footnote{We actually construct the
$p_{n}$s from the $log(x_{n})$ to reduce the influence of extremely
high-density cells that might arise from statistical fluctuations.}.
Figure 1 illustrates the method.

A natural way to quantify the relative dispersion suffered by the
low and high energy components of the burst is to calculate the
Kolmogorov distance (K-dist) between the two constructed pdfs. By
doing so, we are using the less affected (and usually better
sampled) low energy burst profile as a reference for the process of
finding the best cancelation parameter to the dispersion of the
high-energy light-curve, which is more sensitive to the dispersion.

By applying a simultaneous correction as in equation (5) to the
arrival times of each photon in the low and high energy bins, we
want to find the optimum correction $\tau^{*}$ which minimizes the
K-dist between the two bursts:

\begin{equation}\label{equ6}
\tau^{*} : D_{K,\tau^{*}} = \min_{\tau} \sup_{x \in \Re}|F_{X}(x) -
F_{Y}(x)|,
\end{equation}
corresponding to the QG-induced delay. It is important to note that
the cdf is an ideal (and simple) representation to be used for this
purpose of comparing two distributions, which acts like a fitting
measure of the two profiles as the temporal dispersion is
canceled.\footnote{Notice also that we are \textit{not} relying on
confidence intervals and p-values from the Kolmogorov-Smirnov test
to compare the distributions, but rather using the K-dist as a
metric to differentiate them in probability space.}

\begin{figure}[!t]
  \centering
  \includegraphics[width=0.5\textwidth]{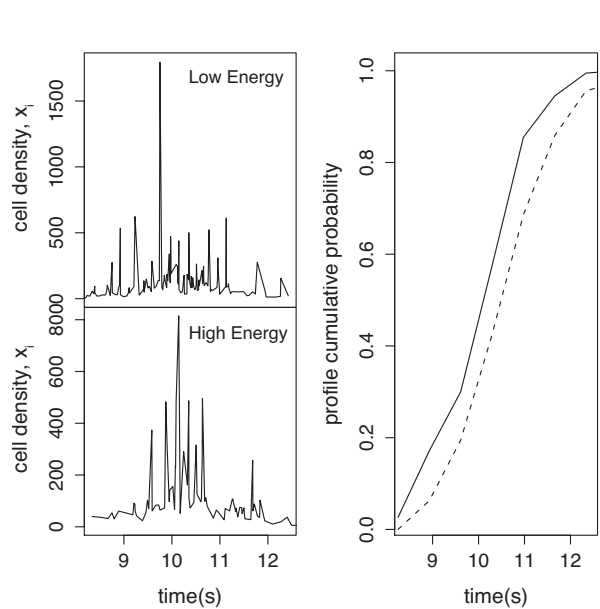}
  \caption{Illustration of the method of Kolmogorov distances for a Gaussian profile.
  The panels to the left represent the low- and high-energy cell density
  representation for the bursts, which appear shifted relative to each other. To the right
is the cdf of their  correspondent probability distributions; the
K-dist is the maximum vertical distance
  between the curves in this plot.}
  \label{kolmogorovfigure}
 \end{figure}

\section{Performance of the Method}
Following \cite{amelino-camelia}, we define a \textit{sensitivity
factor}

\begin{equation}\label{equ4}
\eta \equiv \frac{\Delta t}{\delta t}
\end{equation}
where $\Delta t$ represents the relative delay that two photons of
different energies acquired on their travel from the source as in
equation (2), and $\delta t$ the width of the burst under study.
This parameter measures the power of the method in relation to the
size of the light-curve features, which is the most important factor
in detecting the delays, and by definition will always be calculated
relatively to the delay suffered by a photon of energy equal to the
average energy of all events in the burst.

\begin{figure}[!t]
  \centering
  \includegraphics[width=0.5\textwidth]{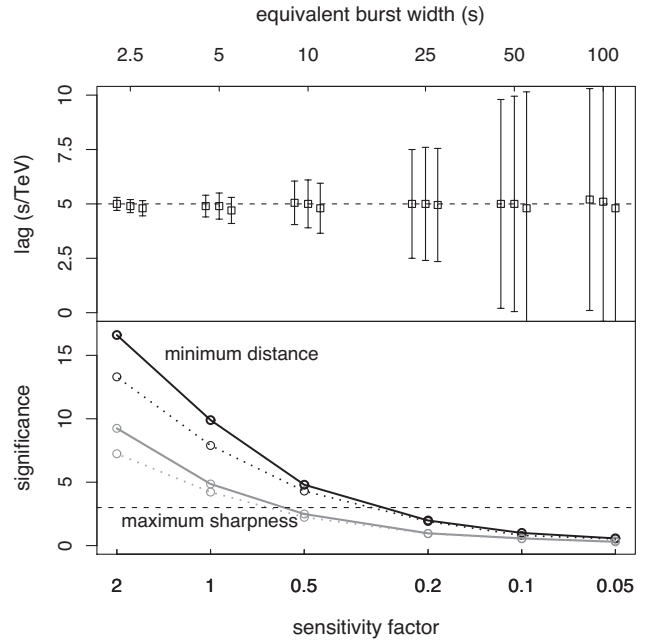}
  \caption{Sensitivity of the Kolmogorov method. The upper panel shows 10000
MC realisations of the method for detecting lags in function of the
sensitivity factor $\eta$. The upper x-axis indicates the equivalent
width of the burst in seconds for a source at 500 Mpc and $\xi=1$,
and each tern of points correspond to 0, 10 and 20\% error in energy
respectively. The bottom panel compares the sensitivity of our
method of minimum distance to that of maximum sharpness methods such
as in \cite{scargle} and \cite{qgmagic}, which have comparable
performances among them.}
  \label{kolmogorovfigure}
 \end{figure}

Typical values for the delays are so small ($\sim$ 1-10 s/TeV) that
in order to detect their effect we have to either rely on the
observation of extremely short bursts, for which $\eta \geq 1$
\cite{bhat}, or on analysis methods sensitive to small deformations
of the burst profile. Apart from GRBs for which $\eta > 0.1\xi$ is
frequent, for AGNs detected in the TeV range, the best case to date
is from the large flare of PKS 2155-304 in 2006 \cite{aharonian}:
its shortest-duration burst of $\sim 2$ min and average photon
energy of $\sim 1$ TeV, imply a sensitivity factor $\eta \sim
0.05\xi$.

The top panel of figure 2 shows the simulated performance of the
K-dist method in function of $\eta$ for a source with spectral index
-2.5 at a distance of 500 Mpc. The sensitivity factor is calculated
assuming $\xi = 1$. Each simulated burst is a MC realisation of an
inhomogeneous poisson process with 1000 events distributed according
to a gaussian rate function. The three adjacent points for each
value of $\eta$ represent respectively 0, 10 or 20\% error in the
energy resolution of the observations, which is introduced in the
simulation at the moment of the correction for the lag. We can see
that the method is only little affected by it, and that the energy
error introduces a small systematic underestimation on the size of
the lag, but always compatible with the true lag within one standard
deviation.

The method as it stands is capable of detecting delays within
3$\sigma$ for $\eta \leq 0.3$ ($\xi = 1$), corresponding to a burst
of 30 s duration for a source at 500 Mpc. Taking a feature
equivalent to the shortest burst in the PKS 2155-304 large flare, we
can probe the presence of QG effects in a scale up to $\sim 3 \times
10^{18}$ GeV, a factor of 10 above the Planck energy, and of the
order of the most accurate upper limits to date on the QG scale.

\begin{figure}[!t]
  \centering
  \includegraphics[width=0.5\textwidth]{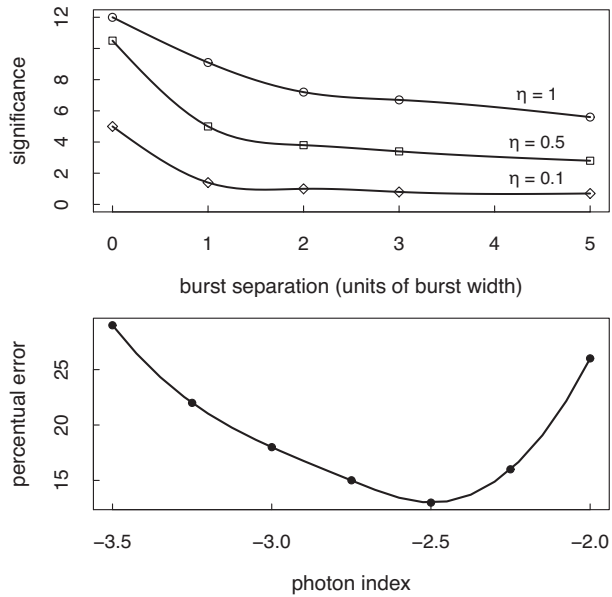}
  \caption{The top panel shows the variations on the performance of
the method for a two-superposed burst configuration; the three
curves correspond to different lag-to-burst width ratios $\eta$. The
bottom panel shows the variation on the sensitivity of detection of
a lag of $\eta = 1$ with the photon index.}
  \label{kolmogorovfigure}
 \end{figure}

Finally, we have tested how the performance of the method depends on
the source spectral index and how it varies in the presence of
multiple overlayed bursts; the results are presented in figure 3. To
test the effect of burst superposition, we generated 4 sets of 10000
MC realisations of a light-curve each, consisting of two identical
gaussian bursts separated by 1, 2, 3 and 5 times the individual
profile widths, respectively. We can see from the upper plot of
figure 3 that the superposition affects the performance of the
method as it broadens the \textit{effective width} of the feature in
which the lag must be detected. The situation is progressively
worsened as the distance between the individual bursts increase,
further spreading the combined profile, until both are completely
separated (around 3 sigma separation), and the curve reaches its
assymtoptic limit. At this point, in practice, the individual bursts
should be treated individually for better results, which should not
represent a problem as long as their separation is comparable or
larger than the average photon lag to be tested; otherwise, there
could have occurred significant "leakage" of photons between the
bursts, which might reflect on the effectiveness of the cancelation
method.

The bottom panel of figure 3 shows the sensitivity curve in function
of the photon index of the burst, which presents an optimal minimum
around an index of -2.5, for a fixed minimum energy threshold. All
bursts generated for testing this effect had the same number of
photons (1000) and the energy boundaries of the low- and high-energy
profiles are chosen so that the average energy difference between
the two profiles and the number of photons in each of them are
maximised. The global minimum of the curve at -2.5 results from the
fact that for very steep or very hard photon indexes both these
factors cannot be ideally optimised.

\section{Conclusions and Possible Variations on the Method}

We have exposed here the general accords of an alternative method to
test for energy dependent lags in HE light-curves. The method draws
inspiration from the unbinned dispersion-cancelation algorithms
independently derived by \cite{scargle} and \cite{qgmagic}, but it
evolves from a maximum sharpness cost-function approach to the
minimization of an appropriate distance metric between low- and
high-energy components of the burst. By doing so we aim to avoid
problems such burst asymmetry that can weaken the assumption of
maximum sharpness. In this regard, the metric minimization approach
has the role of a dynamic fit between the two components of the
profile. We are currently testing ways for further increasing the
sensitivity of the method. The search for new, more appropriate,
distance measures are also under way and are encouraged to be
tested. Applications of this method to VHE and GRB data is underway.

\section*{Acknowledgements}
\noindent U.Barres de Almeida acknowledges a PhD scholarship from
CAPES Foundation, Ministry of Education of Brazil.


\begin{thebibliography}{99}
   \bibitem{qghess} F.~Aharonian, et al. \emph{Phys. Rev. Lett.}
   101, L17 (2009)
   \bibitem{aharonian}     F.~Aharonian, et al. \emph{Astroph.
   Journal} 664, L71 (2007)
   \bibitem{qgmagic}    J.~Albert, et al. \emph{Phys. Lett. B} 668,
   253 (2008)
   \bibitem{albert} J.~Albert, et al. \emph{Phys. Rev. Lett.} 667,
   L21 (2007)
   \bibitem{amelino-camelia}    G.~Amelino-Camelia, et al.
   \emph{Nature} 393, 783 (1998)
   \bibitem{bhat}    C.L.~Bhat, et al. \emph{Nature} 359, 217 (1992)
   \bibitem{ellisa} J.~Ellis, et al. \emph{Astron. Astroph.} 402, 409 (2003)
   \bibitem{ellisb} J.~Ellis, N.E.~Mavromatos and
   D.V.~Nanopoulos \emph{Phys. Lett. B} 665, 412 (2008)
   \bibitem{ellisc} J.~Ellis, N.E.~Mavromatos and D.V.~Nanopoulos
   \emph{Phys. Lett. B} 674, 83 (2009)
   \bibitem{fermi}  Fermi LAT and Fermi GBM Collab., A.~Abdo et al.
   \emph{Science} 323, 1688 (2009)
   \bibitem{feynman}    R.P.~Feynman, R.B.~Leighton and M.~Sands \emph{The
   Feynman Lectures on Physics - Vol. I-31}. Addison-Wesley
   Publishing Co., California (1963)
   \bibitem{fishman}    G.J.~Fishman and C.A.~Meegan \emph{Ann.
   Review Astron. Astroph.} 33, 415 (1995)
   \bibitem{hinshaw}    G.~Hinshaw, et al. \emph{Astroph. Journal Supl.} 180, 225 (2009)
   \bibitem{kolmogorov} A.N.~Kolmogorov \emph{Selected Works of
   Andrei Kolmogorov - Vol.2}, ed. V.M. Tikomirov. Kluwer Academic,
   Boston (1991)
   \bibitem{martinez}   M.~Martinez and M.~Errando \emph{Astrop. Phys.} 31, 226 (2009)
   \bibitem{scargle}    J.D.~Scargle, J.P.~Norris and J.T.~Bonnel
   \emph{Astroph. Journal} 673, 972 (2008)
   \bibitem{wagner} R.~Wagner \emph{arXiv:0901.2932} (2009)
   \bibitem{zolotarev}  V.M.~Zolotarev \emph{Theory Probab. Appl.}
   28, 278 (1983)
 \end{thebibliography}
\end{document}